\documentclass[aps,twocolumn,pra]{revtex4-2}

\usepackage[utf8]{inputenc}
\usepackage[T1]{fontenc}
\usepackage{amsmath}
\usepackage{amssymb}
\usepackage{dsfont}
\usepackage{tikz}
\usepackage[qm]{qcircuit}

\newcommand{\EOp}{\mathsf{E}\kern-1pt\llap{$\vert$}}
\newcommand{\bra}[1]{\langle{#1}|}
\newcommand{\ket}[1]{|{#1}\rangle}

\begin{document}
%%%%%%%%%%%%%%%%%%%%%%%%%%%%%%%%%%%%%%%%%%%%%%%%%%%%%%%%%%%%%%%%%%%%%%%%%%%%%%%%

\title{Description and error analysis of quantum alghorithms in the
  projection evolution model -- the Deutsch algorithm case}

\author{Krzysztof Lider}
\email{lider.krzysztoff@gmail.com}
\affiliation{Institute of Physics, Maria Curie–Skłodowska University,
  pl. Marii Curie-Skłodowskiej 1, 20-031 Lublin, Poland}

\author{Marek Góźdź}
\email{mgozdz@kft.umcs.lublin.pl}
\affiliation{Institute of Computer Science and Mathematics, 
  Maria Curie–Skłodowska University,
  ul. Akademicka 9, 20-033 Lublin, Poland}

\date{\today}

\begin{abstract}
  This work demonstrates that the Deutsch algorithm can be effectively
  modelled using a two-level harmonic oscillator within the second
  quantization formalism. By adopting this framework, evolution
  operators are derived. We present a projection evolution model that
  accurately characterizes the physical state transformation within
  quantum gates. This approach provides a systematic method for finding
  evolution operators, enabling the complete description and prediction
  of state evolution -- including projection errors -- in quantum
  algorithms.
\end{abstract}

\maketitle

%%%%%%%%%%%%%%%%%%%%%%%%%%%%%%%%%%%%%%%%%%%%%%%%%%%%%%%%%%%%%%%%%%%%%%%
\section{Introduction}
%%%%%%%%%%%%%%%%%%%%%%%%%%%%%%%%%%%%%%%%%%%%%%%%%%%%%%%%%%%%%%%%%%%%%%%

Quantum computing may offer a~significant speedup over classical algorithms
primarily due to the quantum parallelism. A foundational example is the
Deutsch algorithm \cite{Deutsch}, which determines the parity of a Boolean
function. As quantum hardware advances, there is an increasing need for robust
physical models to describe the gate dynamics and to predict operational
errors.

Time in quantum mechanics appears in a~parametric form in the solution of the
Schr\"odinger equation. The Pauli principle \cite{Pauli} forbids to construct
the time operator being canonically conjugate the the Hamiltonian. Some
experiments \cite{exp1,exp2,exp3}, however, indicate the need of treating time
as a~quantum observable and not a~classical parameter. A~necessary
reformulation of quantum mechanics leads to the proper inclusion of time in
the model and allows to correctly describe quantum events. It has been shown
in Ref.~\cite{Gozdz} that the projection evolution (PEv) model meets all the
requirements. In this paper we use the PEv approach to describe a~series of
quantum state transformations, the Deutsch algorithm, as evolution of the
input states.

The paper is organized as follows. In Sec.~\ref{sec:PEv} we describe the
projection model of the evolution of a~quantum state. In Sec.~\ref{sec:qGates}
we present that any quantum gate action can be described using the projection
evolution formalism. The Deutsch algorithm is presented in Sec.~\ref{sec:Dalg}
and in Sec.~\ref{sec:DalgE} we discuss this algorithm within the PEv
model. Finally in Sec.~\ref{sec:errors} we apply this method to include error
propagation in the algorithm.

%%%%%%%%%%%%%%%%%%%%%%%%%%%%%%%%%%%%%%%%%%%%%%%%%%%%%%%%%%%%%%%%%%%%%%%
\section{The projection evolution}
\label{sec:PEv}
%%%%%%%%%%%%%%%%%%%%%%%%%%%%%%%%%%%%%%%%%%%%%%%%%%%%%%%%%%%%%%%%%%%%%%%

The projection evolution formalism used in this work is based on a generalized
form of the L\"uders \cite{Luders} type of the projection postulate. We assume
the wave function is dependent on all four spacetime coordinates from which
both time and position are quantum observables. It means that $\psi(t,x)$
represents the probability density of finding the particle in the vicinity of
a~space-time point and that the particle is defined not only in space, but
also along the time axis. The temporal width of the wave function indicates
how wide time segment the particle occupies. It is all a~natural consequence
of treating time as a~quantum observable and the fourth component of the
localization vector. The details of this approach, the construction of the
time operator, and discussion of the results can be found in \cite{Gozdz} and
will be not repeated here. The important issue, however, is that time cannot
serve as the parameter of evolution of a~quantum state. If the wave function
is explicitly $t$-dependent and $t$ is not the classical time, the quantum
evolution must be redefined. In the PEv model the evolution of a~quantum
system is described by a~mapping between Hilbert spaces,
\begin{equation}
  \EOp: {\cal H}_1 \to {\cal H}_2.
\label{eq:EOp}
\end{equation}
We mark the $n$-th evolution step by a~parameter $\tau_n$, which will change
into $\tau_{n+1}$ in the subsequent step of the evolution. It means that the
evolution operator acts on the wave function as follows:
\begin{equation}
  \EOp(\tau_{n+1},\nu_{n+1}) \psi(\tau_n,\nu_n;t,x) = 
  \psi(\tau_{n+1},\nu_{n+1};t,x),
\end{equation}
where $\nu_n$ is the set of quantum numbers at the $n$-th evolution step. If
there is more than one state at the $n+1$ evolution step possible, the
$\nu_{n+1}$ is chosen according to the probability distribution governing the
evolution at this step. For more details on this topic we advice to consult
Ref.~\cite{Gozdz}.

The evolution operators which characterize a given physical process are
formally responsible for quantum evolution of this system. They can have
different forms: unitary operators, POVM, projection operators. In this paper,
the evolution operators represented by an orthogonal resolution of unity are
used:
\begin{eqnarray}
  && \EOp(\tau_n,\nu_n)^\dagger = \EOp(\tau_n,\nu_n), \\
  && \EOp(\tau_n,\nu_n) \ \EOp(\tau_n,\nu'_n) = 
    \delta_{\nu_n\nu'_n} \ \EOp(\tau_n,\nu_n), \\
  && \sum_{\nu_n} \EOp(\tau_n,\nu_n) = \mathds{1},
\end{eqnarray}
where $\mathds{1}$ denotes the unit operator and the sum is taken over
the whole set of possible quantum numbers at the evolution step $n$.

The evolution of a~quantum system described by its density matrices is
given by:
\begin{equation}
  \rho(\tau_{n+1},\nu_{n+1}) = 
  \frac{\EOp(\tau_{n+1},\nu_{n+1}) \rho(\tau_n,\nu_n)
    \EOp(\tau_{n+1},\nu_{n+1})^{\dagger}} 
  {\mathrm{Tr}[\EOp(\tau_{n+1},\nu_{n+1}) \rho(\tau_n,\nu_n) 
    \EOp(\tau_{n+1},\nu_{n+1})^{\dagger}]}.
\end{equation}
The denominator is introduced to properly normalize the density matrix and for
shorter notation we have dropped the indication that $\rho=\rho(t,x)$.

%%%%%%%%%%%%%%%%%%%%%%%%%%%%%%%%%%%%%%%%%%%%%%%%%%%%%%%%%%%%%%%%%%%%%%%
\section{Quantum gates as evolution operators}
\label{sec:qGates}
%%%%%%%%%%%%%%%%%%%%%%%%%%%%%%%%%%%%%%%%%%%%%%%%%%%%%%%%%%%%%%%%%%%%%%%

In quantum information theory a~qubit can be represented by a~vector in
a~two-dimensional complex Hilbert space ${\cal H}$. A~one-qubit quantum gate
is defined as a~unitary operation:
\begin{equation}
  \mathrm{G}: {\cal H} \to {\cal H}
\end{equation}
and, after choosing the vector basis, can be represented by a~$2\times 2$
unitary matrix. Similarly, an~$n$-qubit quantum gate is given by a~unitary
mapping:
\begin{equation}
  \mathrm{G}_n: {\cal H}^n \to {\cal H}^n
\end{equation}
and represented by a~$2^n\times 2^n$ unitary matrix.

It is clearly visible that the quantum gate acts in the same way as the
evolution operator (\ref{eq:EOp}). Below we construct the evolution operator
for a~generic $n$-qubit gate.

Let the gate be given in its matrix form by:
\begin{equation}
  \mathrm{G}_n = 
  \left( \begin{matrix}
    A_{11} & A_{12} & \cdots & A_{1N} \\
    A_{21} & A_{22} & \cdots & A_{2N} \\
    \vdots & \vdots & \ddots & \vdots \\
    A_{N1} & A_{N2} & \cdots & A_{NN}
  \end{matrix} \right)
\label{eq:Gn}
\end{equation}
where $N=2^n$ and $\mathrm{G}_n^\dagger = \mathrm{G}_n^{-1}$. For the qubit
$\ket{\phi} = (x_1, x_2,\dots x_N)^T$ one gets
\begin{equation}
\mathrm{G}_n \ket{\phi} =
  \left( \begin{matrix}
    A_{11}x_{1} + A_{12}x_{2} + \cdots + A_{1n}x_{n} \\
    A_{21}x_{1} + A_{22}x_{2} + \cdots + A_{2n}x_{n} \\
    \vdots \\
    A_{n1}x_{1} + A_{n2}x_{2} + \cdots + A_{nn}x_{n}
  \end{matrix} \right).
\end{equation}

The evolution operator $\EOp_{\mathrm{G}_n}$ 
\begin{equation}
  \EOp_{\mathrm{G}_n} = 
  \left( \begin{matrix}
    a_{11} & a_{12} & \cdots & a_{1n} \\
    a_{21} & a_{22} & \cdots & a_{2n} \\
    \vdots & \vdots & \ddots & \vdots \\
    a_{n1} & a_{n2} & \cdots & a_{nn}
  \end{matrix} \right)
\end{equation}
acts on the density matrix
\begin{equation}
  \rho_{in}=\ket{\phi}\bra{\phi} = 
  \left( \begin{matrix}
    x_1 x_1^* & x_1 x_2^* & \cdots & x_1 x_N^* \\
    x_2 x_1^* & x_2 x_2^* & \cdots & x_2 x_N^* \\
    \vdots & \vdots & \ddots & \vdots \\
    x_N x_1^* & x_N x_2^* & \cdots & x_N x_N^*
  \end{matrix} \right)
\end{equation}
resulting in
\begin{equation}
  \rho_{out} = \frac{\EOp_{\mathrm{G}_n} \rho_{in} \EOp_{\mathrm{G}_n}^\dagger}
  {\mathrm{Tr}(\EOp_{\mathrm{G}_n} \rho_{in} \EOp_{\mathrm{G}_n}^\dagger)}.
\label{eq:rhoOut}
\end{equation}
Since quantum gates are unitary by assumption, the denominator in
Eq.~(\ref{eq:rhoOut}) drops out,
$\mathrm{Tr}(\EOp_{\mathrm{G}_n} \rho_{in} \EOp_{\mathrm{G}_n}^\dagger)
= 1$, which leads to
\begin{equation}
  \EOp_{\mathrm{G}_n} \rho_{in} \EOp_{\mathrm{G}_n}^\dagger = 
  \rho_{out} = 
  \mathrm{G}_n \rho_{in} \mathrm{G}_n^{\dagger}.
\end{equation}
The solution is $|a_{kl}|^2 = |A_{kl}|^{2}$ or equivalently
$a_{kl}=A_{kl} e^{i\alpha_{kl}}$. Again, due to the unitarity of the
quantum gate, all the phases have to have the same value, which can be
checked by a~direct calculation. We finish therefore with the simple
solution:
\begin{equation}
  a_{kl} = A_{kl} e^{i\alpha}.
\end{equation}
We conclude that the projection evolution operator for a~quantum gate
given by Eq.~(\ref{eq:Gn}) is:
\begin{equation}
  \EOp_{\mathrm{G}_n} = e^{i\alpha} \left( \begin{matrix}
    A_{11} & A_{12} & \cdots & A_{1N} \\
    A_{21} & A_{22} & \cdots & A_{2N} \\
    \vdots & \vdots & \ddots & \vdots \\
    A_{N1} & A_{N2} & \cdots & A_{NN}
  \end{matrix} \right).
\end{equation} 

There are two types of gates in the Deutsch algorithm. The Hadamard gate
will be represented by the evolution operator
\begin{equation}
  \EOp_H = \frac{e^{i \alpha}}{\sqrt{2}}
  \left( \begin{matrix} 1 & 1 \\ 1 & -1 \end{matrix} \right).
\label{eq:EOp.H}
\end{equation}
For a~classical function $f: \{0,1\} \to \{0,1\}$ the oracle quantum
gate is defined as:
\begin{equation}
  U_f \ket{x} \ket{y} = \ket{x} \ket{y \oplus f(x)}.
\label{eq:defUf}
\end{equation}
The evolution operators for $U_f$ are presented in Tab.~\ref{tab:Uf}

%%%%%%%%%%%%%%%%%%%%%%%%%%%%%%%%%%%%%%%%%%%%%%%%%%%%%%%%%%%%%%%%%%%%%%%%
\begin{table}[ht!]
  \caption{Projection evolution operators of the oracle gate for all
    four $f$ functions.}
  \label{tab:Uf}
  \begin{tabular}{|c|c|}
    \hline
    $f$ function & $\EOp$ operator \\ 
    \hline\hline
    $f_1(0)=0$, $f_1(1)=0$ &
    $\EOp_1 = e^{i \alpha}
    \begin{pmatrix} 1 & 0 & 0 & 0 \\
                    0 & 1 & 0 & 0 \\
                    0 & 0 & 1 & 0 \\
                    0 & 0 & 0 & 1 \end{pmatrix}$ \\
    \hline
    $f_2(0)=0$, $f_2(1)=1$ & 
    $\EOp_2 = e^{i \alpha} 
    \begin{pmatrix} 1 & 0 & 0 & 0 \\ 
                    0 & 1 & 0 & 0 \\ 
                    0 & 0 & 0 & 1 \\ 
                    0 & 0 & 1 & 0 \end{pmatrix}$ \\
    \hline
    $f_3(0)=1$, $f_3(1)=0$ & 
    $\EOp_3 = e^{i \alpha} 
    \begin{pmatrix} 0 & 1 & 0 & 0 \\ 
                    1 & 0 & 0 & 0 \\ 
                    0 & 0 & 1 & 0 \\ 
                    0 & 0 & 0 & 1 \end{pmatrix}$ \\ 
    \hline
    $f_4(0)=1$, $f_4(1)=1$ & 
    $\EOp_4 = e^{i \alpha} 
    \begin{pmatrix} 0 & 1 & 0 & 0 \\ 
                    1 & 0 & 0 & 0 \\ 
                    0 & 0 & 0 & 1 \\ 
                    0 & 0 & 1 & 0 \end{pmatrix}$ \\   
    \hline
  \end{tabular}
\end{table}
%%%%%%%%%%%%%%%%%%%%%%%%%%%%%%%%%%%%%%%%%%%%%%%%%%%%%%%%%%%%%%%%%%%%%%%%

%%%%%%%%%%%%%%%%%%%%%%%%%%%%%%%%%%%%%%%%%%%%%%%%%%%%%%%%%%%%%%%%%%%%%%%
\section{The Deutsch algorithm}
\label{sec:Dalg}
%%%%%%%%%%%%%%%%%%%%%%%%%%%%%%%%%%%%%%%%%%%%%%%%%%%%%%%%%%%%%%%%%%%%%%%

Having an unknown function $f: \{0,1\} \to \{0,1\}$ one wants to know if it is
constant or balanced. To answer this question a~classical algorithm needs to
compute this function for both of its arguments. The quantum algorithm
proposed by D.Deutsch \cite{Deutsch} can solve this problem two times faster,
computing the required values $f(0)$ and $f(1)$ in one go. This was the first
presentation of the quantum parallelism, which showed the possible higher
efficiency of quantum algorithms over classical ones.

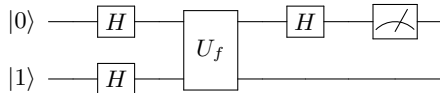
\begin{figure}[b]
  \centerline{
    \Qcircuit @C=1em @R=1em {
      & \lstick{\ket{0}} & {} \qw & \gate{H} & \qw & \multigate{1}{U_f} 
      & \qw & \gate{H} & \qw & \meter & \qw \\
      & \lstick{\ket{1}} & {} \qw & \gate{H} & \qw & \ghost{U_f} 
      & \qw & \qw & \qw & \qw & \qw}
  }
  \caption{The circuit of the Deutsch algorithm.}
  \label{fig:Dcirc}
\end{figure}

The Deutsch algorithm scheme is presented in Fig.~\ref{fig:Dcirc}. It
consists of two registers, three Hadamard gates and one oracle
gate. The input state is $\ket{0}\ket{1}$ and the first register is
measured at the end. 

%%%%%%%%%%%%%%%%%%%%%%%%%%%%%%%%%%%%%%%%%%%%%%%%%%%%%%%%%%%%%%%%%%%%%%%%
%%%%%%%%%%%%%%%%%%%%%%%%%%%%%%%%%%%%%%%%%%%%%%%%%%%%%%%%%%%%%%%%%%%%%%%%

The oracle gate is defined by Eq.~(\ref{eq:defUf}); it is reversing the second
qubit if $f(x)=1$ while for $f(x)=0$ the gate copies its input on the
output. This allows to rewrite the definition of $U_f$ using the Kronecker
delta symbols:
\begin{equation}
U_f \ket{x} \ket{y} = \delta_{0 f(x)} \ket{x} \ket{y} + 
                      \delta_{1 f(x)} (\mathds{1} \otimes \mathrm{NOT}) 
                      \ket{x} \ket{y}.
\end{equation}

It is straightforward to check that the output register has the form
\begin{equation}
  \frac{1}{2\sqrt{2}}(A\ket{0}+B\ket{1})(\ket{0}-\ket{1}), \label{eq:AB}
\end{equation}
where
$A=\delta_{0 f(0)}-\delta_{1 f(0)}+\delta_{0 f(1)}-\delta_{1 f(1)}$ and
$B=\delta_{0 f(0)}-\delta_{1 f(0)}-\delta_{0 f(1)}+\delta_{1 f(1)}$. If
$f$ is constant, $f(0)=f(1)$, then $A=\pm 2, B=0$ and the measurement of
the first qubit will always result in $\ket{0}$. For balanced function,
$f(0) \neq f(1)$, $A=0, B=\pm 2$ and the measurement will always give
$\ket{1}$.

%%%%%%%%%%%%%%%%%%%%%%%%%%%%%%%%%%%%%%%%%%%%%%%%%%%%%%%%%%%%%%%%%%%%%%%
\section{The Deutsch algorithm as the evolution of a~quantum state}
\label{sec:DalgE}
%%%%%%%%%%%%%%%%%%%%%%%%%%%%%%%%%%%%%%%%%%%%%%%%%%%%%%%%%%%%%%%%%%%%%%%

Any quantum algorithm is in fact a~series of transformations of
a~quantum state. From the point of view of quantum mechanics, to
describe such a~process one needs the state on which the qubit is
encoded and perform the evolution of this state according to the quantum
gates and operations used in the algorithm. In what follows we rewrite
the Deutsch algorithm using quantum harmonic oscillator states and the
projection evolution model.

The harmonic oscillator is defined by the number $n$ representing the $n$-th
excited state $\ket{n}$. The ground state is then given by $\ket{0}$,
$\ket{1}$ denotes the first excited state and so on. In what follows we
represent the one-qubit computational basis by the ground state:
\begin{equation}
  \ket{0} = \left (\frac{m \omega}{\pi \hbar} \right)^{\frac{1}{4}} 
  e^{\frac{-m \omega x^2}{2 \hbar}}
%  H_0 \left(\sqrt{\frac{m \omega}{\hbar}}x \right) 
  \begin{pmatrix} 1 \\ 0 \end{pmatrix}
\end{equation}
% where the Hermite polynomial $H_0(\sqrt{\frac{m \omega}{\hbar}}x) = 1$
and the first excited state:
\begin{equation}
  \ket{1} = \left (\frac{m \omega}{\pi \hbar} \right)^{\frac{1}{4}} 
  e^{\frac{-m \omega x^2}{2 \hbar}} 
  H_1 \left(\sqrt{\frac{m \omega}{\hbar}}x \right) 
  \begin{pmatrix} 0 \\ 1 \end{pmatrix}
\end{equation}
where the Hermite polynomial $H_1(\sqrt{\frac{m\omega}{\hbar}}x)=2x$.

In the second quantization formalism one introduces the annihilation
operator $\hat{a}$ and the creation operator $\hat{a}^{\dagger}$ defined
by the relations:
\begin{eqnarray}
  && \hat{a} \ket{n} = \sqrt{n} \ket{n-1}, \\ 
  && \hat{a}^{\dagger} \ket{n} = \sqrt{n+1} \ket{n+1},
\end{eqnarray}
therefore any excited state of the harmonic oscillator can be obtained
from the ground state by the repeated action of the creation operator,
\begin{align}
  \ket{n}=\frac{(\hat{a}^{\dagger})^{n}}{\sqrt{n!}} \ket{0}.
\end{align}

In order to describe the Deutsch algorithm we label the evolution steps by
$\tau_0,\dots,\tau_4$ and recall that $\tau_i$ is not time but the parameter
which enumerates subsequent steps. In general, the state is projected onto the
time axis during the $\tau_0$ and $\tau_4$ steps only, i.e., at the beginning
of the algorithm and at the final measurement. The quantum gates do not need
to localize the qubits in time.

%%%%%%%%%%%%%%%%%%%%%%%%%%%%%%%%%%%%%%%%%%%%%%%%%%%%%%%%%%%%%%%%%%%%%%%%
\begin{figure}
\centering
\tikzset{every node/.style={inner sep=0}}
\tikz[baseline=(qc.base)]{
% --- oryginalny schemat ---
  \node (qc) {
    $\Qcircuit @C=1em @R=1em {
      & \lstick{\ket{0}} & {} \qw & \gate{H} 
      & \qw & \multigate{1}{U_f} & \qw & \gate{H} 
      & \qw & \meter & \qw & {} \qw \\
      & \lstick{\ket{1}} & {} \qw & \gate{H} 
      & \qw & \ghost{U_f} & \qw & \qw 
      & \qw & \qw & \qw & {} \qw
    }$
  };

  \begin{scope}[yshift=-15em] % większy yshift = niżej
  \foreach \x/\lab in 
  {2.0/\tau_{0},3.7/\tau_1,7.5/\tau_2,11.6/\tau_3,15.2/\tau_4} 
  {\draw[dashed, very thin] 
    ([xshift=\x em, yshift=+8mm]qc.west) -- ++(0,-17.7mm) 
    node[below, xshift=0.2em, yshift=-0.4mm] {$\lab$};
  }

% oś czasu t
  \draw[->, thick] ([xshift=1.0em,yshift=-9mm]qc.west)
      -- ([xshift=18.3em,yshift=-9mm]qc.west) node[right] {$\tau$};
% oś położeń x
%  \draw[->, thick] ([xshift=1.0em,yshift=-15mm]qc.west)
%      -- ([xshift=18.3em,yshift=-15mm]qc.west) node[right] {$x$};
% oznaczenia t_0 i t_7
%  \node[below] at ([xshift=2.2em,yshift=-10mm]qc.west) {$\tau_0$};
%  \node[below] at ([xshift=5.7em,yshift=-10mm]qc.west) {$\tau_1$};
%  \node[below] at ([xshift=10em,yshift=-10mm]qc.west) {$\tau_2$};
%  \node[below] at ([xshift=13.4em,yshift=-10mm]qc.west) {$\tau_3$};
%  \node[below] at ([xshift=17.4em,yshift=-10mm]qc.west) {$\tau_4$};
  \end{scope}
}
\caption{The evolution steps in the Deutsch algorithm.}
\label{fig:Dcirc2}
\end{figure}
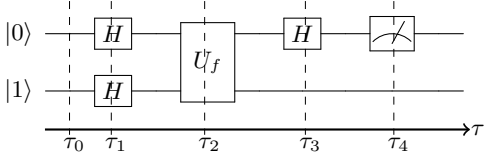
%%%%%%%%%%%%%%%%%%%%%%%%%%%%%%%%%%%%%%%%%%%%%%%%%%%%%%%%%%%%%%%%%%%%%%%%

Let us denote by $\ket{\psi_1}$ the first and by $\ket{\psi_2}$ the
second qubit, so that the state of the register at the evolution step
$\tau$ is
\begin{equation}
  \ket{\psi_1(\tau)} \otimes \ket{\psi_2(\tau)}.
\end{equation}
The input state at $\tau_0$ has the form
\begin{equation}
  \ket{\psi_1(\tau_0)} \otimes \ket{\psi_2(\tau_0)} = 
  \ket{0} \otimes \ket{1} = 
  \begin{pmatrix} 1 \\ 0 \end{pmatrix} \otimes
  \begin{pmatrix} 0 \\ 1 \end{pmatrix}.
\end{equation}

At $\tau_1$ the first qubit is transformed as
$\ket{\psi_1(\tau_1)} = H \ket{0} = \frac{1}{\sqrt{2}}
(\ket{0}+\ket{1})$ with its density matrix given by
\begin{equation}
  \rho_1(\tau_1) = \ket{\psi_1(\tau_1)} \bra{\psi_1(\tau_1)} = 
  \frac{1}{2} \begin{pmatrix} 1 & 1 \\ 1 & 1 \end{pmatrix}. 
\label{eq:rho1tau1}
\end{equation}

Without loss of generality we may set $\alpha=0$ in
Eq.~(\ref{eq:EOp.H}). The projection evolution operator for the Hadamard
gate in the second quantization takes the form
\begin{equation}
  \EOp_H = \frac{1}{\sqrt{2}} 
  \begin{pmatrix} a^{\dagger} & -a^{\dagger} \\ a & a \end{pmatrix},
\label{eq:EopH}
\end{equation}
which can be checked by direct comparison with Eq.~(\ref{eq:rho1tau1}):
\begin{eqnarray}
  \EOp_H \rho_1(\tau_0) \EOp_H^\dagger &=& 
  \frac{1}{\sqrt{2}} 
  \begin{pmatrix} a^{\dagger} & -a^{\dagger} \\ a & a \end{pmatrix} 
  \begin{pmatrix} 1 & 0 \\ 0 & 0 \end{pmatrix} 
  \frac{1}{\sqrt{2}} 
  \begin{pmatrix} a & a^{\dagger} \\ -a & a^{\dagger} \end{pmatrix}
 \nonumber \\
  &=& 
  \frac{1}{2} \begin{pmatrix} 1 & 1 \\ 1 & 1 \end{pmatrix}.
\end{eqnarray}

The same operation is performed on the second qubit,
$\ket{\psi_2(\tau_1)} = H \ket{1} = \frac{1}{\sqrt{2}}
(\ket{0}-\ket{1})$, leading to the density matrix
\begin{equation}
  \rho_2(\tau_1) = \ket{\psi_2(\tau_1)} \bra{\psi_2(\tau_1)} =
  \frac{1}{2} \begin{pmatrix} 1 & -1 \\ -1 & 1 \end{pmatrix}. 
\label{eq:rho2tau1}
\end{equation}

At the evolution step $\tau_2$ the qubits enter the oracle gate $U_f$. The
resulting density matrices written in the two-qubit basis can be found in
Tab.~\ref{tab:EUf}. We notice that the output from the $U_f$ gate is the same
for $f$ being a~constant function ($f_1$ and $f_4$) but different for $f$
being balanced ($f_2$ and $f_3$), which is crucial for the Deutsch algorithm
to work.

%%%%%%%%%%%%%%%%%%%%%%%%%%%%%%%%%%%%%%%%%%%%%%%%%%%%%%%%%%%%%%%%%%%%%%%%
\begin{table}[ht!]
  \caption{Density matrices on the output of the oracle gate for
    different $f$ functions.}
  \label{tab:EUf}
  \begin{tabular}{|c|c|}
    \hline
    $f$ function & $\rho(\tau_2)$ \\ 
    \hline\hline
    $f_1(0)=0$, $f_1(1)=0$ &
    $\EOp_1 \rho(\tau_1) \EOp_1^\dagger = \frac{1}{4}
    \begin{pmatrix} 1 & -1 &  1 & -1 \\ 
                   -1 &  1 & -1 &  1 \\ 
                    1 & -1 &  1 & -1 \\ 
                   -1 &  1 & -1 &  1 \end{pmatrix}$ \\
    \hline
    $f_2(0)=0$, $f_2(1)=1$ & 
    $\EOp_2 \rho(\tau_1) \EOp_2^\dagger = \frac{1}{4} 
    \begin{pmatrix} 1 & -1 & -1 &  1 \\ 
                   -1 &  1 &  1 & -1 \\ 
                   -1 &  1 &  1 & -1 \\ 
                    1 & -1 & -1 &  1 \end{pmatrix}$ \\
    \hline
    $f_3(0)=1$, $f_3(1)=0$ & 
    $\EOp_3 \rho(\tau_1) \EOp_3^\dagger = \frac{1}{4}
    \begin{pmatrix} 1 & -1 & -1 &  1 \\ 
                   -1 &  1 &  1 & -1 \\ 
                   -1 &  1 &  1 & -1 \\ 
                    1 & -1 & -1 &  1 \end{pmatrix}$ \\ 
    \hline
    $f_4(0)=1$, $f_4(1)=1$ & 
    $\EOp_4 \rho(\tau_1) \EOp_4^\dagger = \frac{1}{4}
    \begin{pmatrix} 1 & -1 &  1 & -1 \\ 
                   -1 &  1 & -1 &  1 \\ 
                    1 & -1 &  1 & -1 \\ 
                   -1 &  1 & -1 &  1 \end{pmatrix}$ \\
    \hline
  \end{tabular}
\end{table}
%%%%%%%%%%%%%%%%%%%%%%%%%%%%%%%%%%%%%%%%%%%%%%%%%%%%%%%%%%%%%%%%%%%%%%%%

The transformation at the evolution step $\tau_3$ is to apply a~Hadamard
gate (\ref{eq:EopH}) to the first qubit, $H \otimes \mathds{1}$. The
relevant evolution operator is given by
\begin{equation}
  \EOp_{H\otimes \mathds{1}} = \frac{1}{\sqrt{2}}
  \begin{pmatrix}
    a^{\dagger} & 0 & -a^{\dagger} & 0 \\
    0 & a^{\dagger} & 0 & -a^{\dagger} \\
    a & 0 & -a & 0 \\
    0 & a & 0 & -a
  \end{pmatrix}
\end{equation}
and the density matrix can be obtained by computing
\begin{equation}
  \rho(\tau_3) = 
  \EOp_{H\otimes \mathds{1}} \rho(\tau_2) \EOp_{H\otimes \mathds{1}}^\dagger.
\end{equation}
Based on the state given in Eq.~\eqref{eq:AB}, the $\rho(\tau_3)$ matrix can
be written as:
\begin{eqnarray}
  \rho(\tau_3) = \frac{1}{8} 
  \begin{pmatrix}
    |A|^2 & -|A|^2 & AB^* & -AB^* \\
    -|A|^2 & |A|^2 & -AB^* & AB^* \\
    BA^* & -BA^* & |B|^2 & -|B|^2 \\
    -BA^* & BA^* & -|B|^2 & |B|^2
  \end{pmatrix}.
\label{eq:AB2}
\end{eqnarray}
The values for different $f$ functions are presented in Tab.~\ref{tab:tau3}.

%%%%%%%%%%%%%%%%%%%%%%%%%%%%%%%%%%%%%%%%%%%%%%%%%%%%%%%%%%%%%%%%%%%%%%%%
\begin{table}[ht!]
  \caption{The state of the register at the evolution step $\tau_3$.}
  \label{tab:tau3}
  \begin{tabular}{|c|c|c|}
    \hline
    $f$ function & coefficients & $\rho(\tau_3)$ \\ 
    \hline\hline
    $f_1(0)=0$, $f_1(1)=0$ & $|A|=2$, $B=0$ &
    $
    \frac{1}{2} \begin{pmatrix}
      1 & -1 & 0 & 0 \\
      -1 & 1 & 0 & 0 \\
      0 & 0 & 0 & 0 \\
      0 & 0 & 0 & 0
    \end{pmatrix} $ \\
    \hline
    $f_2(0)=0$, $f_2(1)=1$ & $A=0$, $|B|=2$ &
    $
    \frac{1}{2} \begin{pmatrix}
      0 & 0 & 0 & 0 \\
      0 & 0 & 0 & 0 \\
      0 & 0 & 1 & -1 \\
      0 & 0 & -1 & 1
    \end{pmatrix} $ \\
    \hline
    $f_3(0)=x$, $f_3(1)=0$ & $A=0$, $|B|=2$ & 
    $
    \frac{1}{2} \begin{pmatrix}
      0 & 0 & 0 & 0 \\
      0 & 0 & 0 & 0 \\
      0 & 0 & 1 & -1 \\
      0 & 0 & -1 & 1
    \end{pmatrix} $ \\ 
    \hline
    $f_4(0)=1$, $f_4(1)=1$ & $|A|=2$, $B=0$ & 
    $
    \frac{1}{2} \begin{pmatrix}
      1 & -1 & 0 & 0 \\
      -1 & 1 & 0 & 0 \\
      0 & 0 & 0 & 0 \\
      0 & 0 & 0 & 0
    \end{pmatrix} $ \\
    \hline
  \end{tabular}
\end{table}
%%%%%%%%%%%%%%%%%%%%%%%%%%%%%%%%%%%%%%%%%%%%%%%%%%%%%%%%%%%%%%%%%%%%%%%%

Finally, at the step $\tau_4$, the first qubit is measured, which means that
it is projected onto the computational basis states,
\begin{equation}
\EOp_M=\ket{0}\bra{0}+\ket{1}\bra{1}.
\end{equation}
In general the measurements are special stages of the quantum evolution
because they are irreversible and result in the collapse of the wave
function. In the case of the Deutsch algorithm the final measured state is
$\ket{0}$ or $\ket{1}$, indicating the type of the $f$ function, as can be
clearly seen from Tab.~\ref{tab:tau3}.

%%%%%%%%%%%%%%%%%%%%%%%%%%%%%%%%%%%%%%%%%%%%%%%%%%%%%%%%%%%%%%%%%%%%%%%
\section{Error propagation}
\label{sec:errors}
%%%%%%%%%%%%%%%%%%%%%%%%%%%%%%%%%%%%%%%%%%%%%%%%%%%%%%%%%%%%%%%%%%%%%%%

Quantum systems are not immune to errors. By its nature, the quantum state
undergoes spontaneous decoherence and its stabilization over resonable times
is of utmost importance for quantum computing. The three main cathegories of
errors are the bit-flip $\ket{0} \leftrightarrows \ket{1}$, the phase error
$\ket{x} \to e^{i\alpha}\ket{x}$, and entanglement. In this section we discuss
the possibility, that the Hadamard gates have the chance to perform a~bit flip
on the input qubit. We assume for simplicity, that the $U_f$ gate, which
realizes a~classical function $f$, is error-free.

A~quantum gate may behave in many different ways, from which the two
main cathegories are the projection and unitary gates. The {\it
  projection gate} computes the result and performs a~measurement; the
measured value is then sent to its output. For example, if the correct
behaviour of gate $G$ is:
\begin{equation}
  G\ket{x} = \ket{0},
\end{equation}
the projection version od $G$ works as follows:
\begin{eqnarray}
  G\ket{x} &=&  a \ket{0} + \sqrt{1-|a|^2} \ket{1} \\
  && \xrightarrow{\text{projection}}
  \left\{ \begin{array}{ll}
            \ket{0} & \text{with probability } |a|^2 \\
            \ket{1} & \text{with probability } 1-|a|^2
          \end{array}
        \right.
  \nonumber
\end{eqnarray}

If $G$ were a~{\it unitary gate}, the output will be the quantum superposition
of both results,
\begin{equation}
  G\ket{x} =  a \ket{0} + \sqrt{1-|a|^2} \ket{1}.
\end{equation}

In the case of the bit-flip error, both types of gates lead to the same final
result, as each algorithm has a~measurement at the end and it makes no
difference if one measures the final state or performs a~series of
measurements during the computation. The phase error, however, is canceled by
the projection gates, while in the case of unitary gates it is transmitted to
the subsequent gates and may alter the results.

We assume that the three Hadamard gates can incorrectly read their input
and are of the unitary type. The first gate at the first line $H_1$ has
probability $|\alpha|^2$ of producing the correct and $|\beta|^2$ of
incorrect result, with $|\alpha|^2 + |\beta|^2=1$. The first Hadamard
gate on the second line $H_2$ is characterized by probabilities
$|\gamma|^2$ and $|\delta|^2$, with $|\gamma|^2 + |\delta|^2=1$, and the
second gate at the first line $H_3$ by $|\mu|^2$ and $|\kappa|^2$, with
$|\mu|^2+|\kappa|^2=1$.

We restrict the discussion to the $f_1$ function only, $f_1(0)=f_1(1)=0$, as
other cases are analogous. The register at the step $\tau_2$ now reads:
\begin{eqnarray}
  && \ket{\psi(\tau_2)} \\ 
  && = 
  \left[
    \alpha \frac{\ket{0}+\ket{1}}{\sqrt{2}} + 
    \beta  \frac{\ket{0}-\ket{1}}{\sqrt{2}} \right]
  \left[
    \gamma \frac{\ket{0}-\ket{1}}{\sqrt{2}} + 
    \delta \frac{\ket{0}+\ket{1}}{\sqrt{2}} \right].
  \nonumber
\end{eqnarray}

The gate $U_f$ for our choice of $f$ does not change the register, and the last
Hadamard gate $H_3$ changes the first line as 
\begin{eqnarray}
  \psi_1(\tau_3) &=& H_3 \left[
    \alpha \frac{\ket{0}+\ket{1}}{\sqrt{2}} + 
    \beta  \frac{\ket{0}-\ket{1}}{\sqrt{2}} \right] \nonumber \\
  &=& 
  N \left[
  \alpha (\mu + \kappa) \ket{0} + \beta (\mu - \kappa) \ket{1}
  \right],
\end{eqnarray}
where $N$ is the normalization constant:
\begin{equation}
  N = \left[ 1 + 2\mu\kappa(\alpha^2 - \beta^2) \right]^{-1/2}.
\end{equation}
In the special case in which both gates have identical performance,
$\alpha=\mu$ and $\beta=\kappa=\sqrt{1-\alpha^2}$ we get the
normalization constant
\begin{equation}
  N = \left[ 1 + 2\alpha\sqrt{1-\alpha^2} (2\alpha^2 - 1) \right]^{-1/2}
\end{equation}
and the measurement probabilities
\begin{eqnarray}
  && \mathrm{Prob}(\ket{0}) = \frac
  {\alpha^2 (\alpha + \sqrt{1-\alpha^2})^2}
  {1 + 2\alpha\sqrt{1-\alpha^2}(2\alpha^2-1)}, \\
  && \mathrm{Prob}(\ket{1}) = \frac
  {(\alpha^2-1) (2\alpha\sqrt{1-\alpha^2}-1)}
  {1 + 2\alpha\sqrt{1-\alpha^2}(2\alpha^2-1)}.
\end{eqnarray}

\begin{figure}
  \centering 
  \includegraphics[width=0.8\columnwidth]{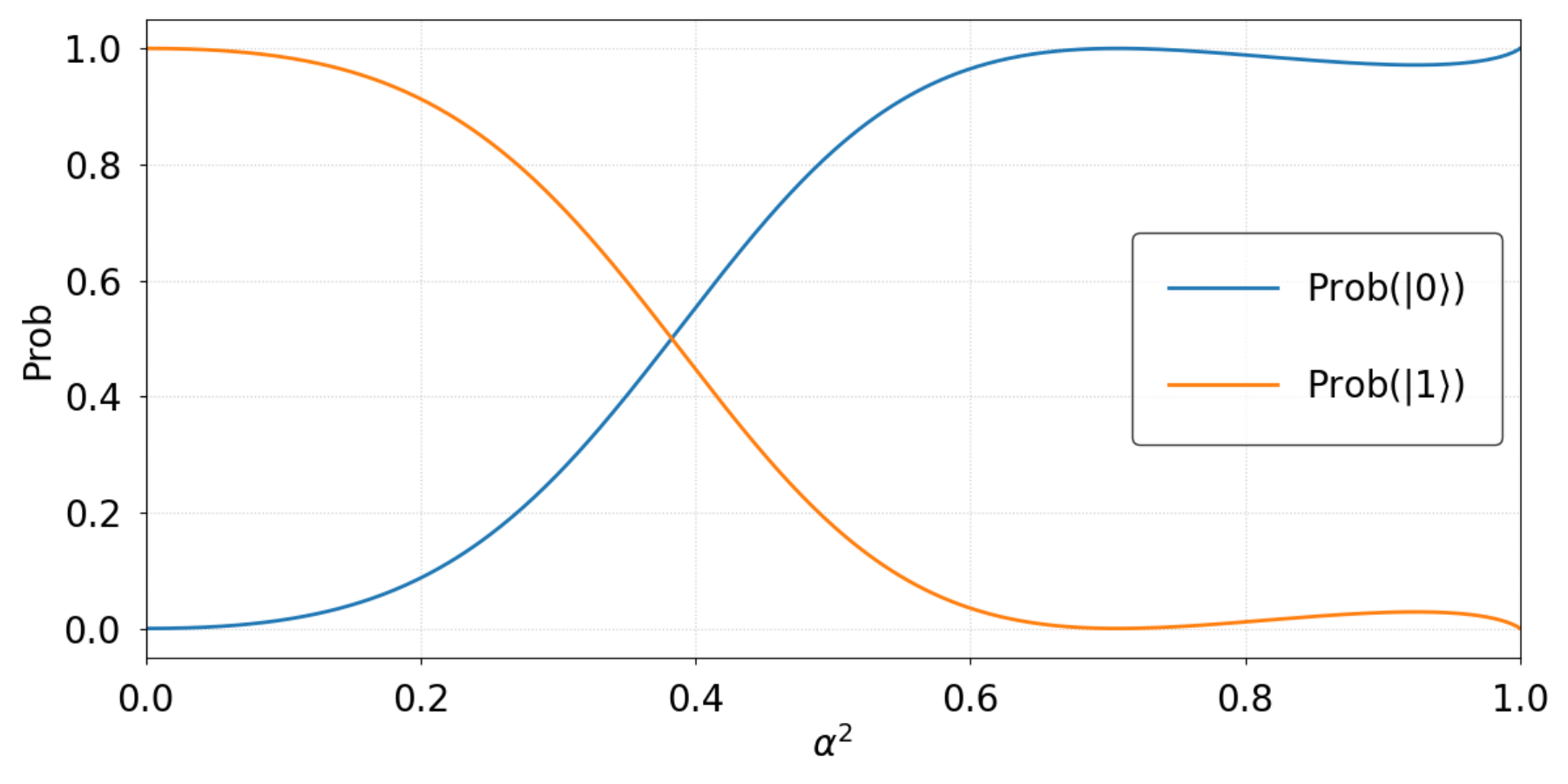}
  \caption{The probability of the correct result,
    $\text{Prob}(\ket{0})$, and incorrect result,
    $\text{Prob}(\ket{1})$, are functions of the probability of the
    correct single Hadamard gate result $\alpha^2$.}
  \label{fig2}
\end{figure}

For $f_1(0) = f_1(1) = 0$, the function $\mathrm{Prob}(\ket{0})$ denotes the
probability that the algorithm will produce the correct result, while
$\mathrm{Prob}(\ket{1})$ denotes the probability that it will produce the
incorrect result. These functions are presented as a function of the
probability of a correct single Hadamard gate result $\alpha^2$ in
Fig.~\ref{fig2}. Within the displayed range of possible values of $\alpha^2$,
the probability $\mathrm{Prob}(\ket{0})$ has a local maximum, which does not
exist for the algorithm with a single Hadamard gate at the first line. This
effect arises because in the case of two Hadamard gates, it is possible for
both gates to make errors, in which case the final result of the algorithm is
stil correct.

\begin{figure}
  \centering
  \includegraphics[width=0.8\columnwidth]{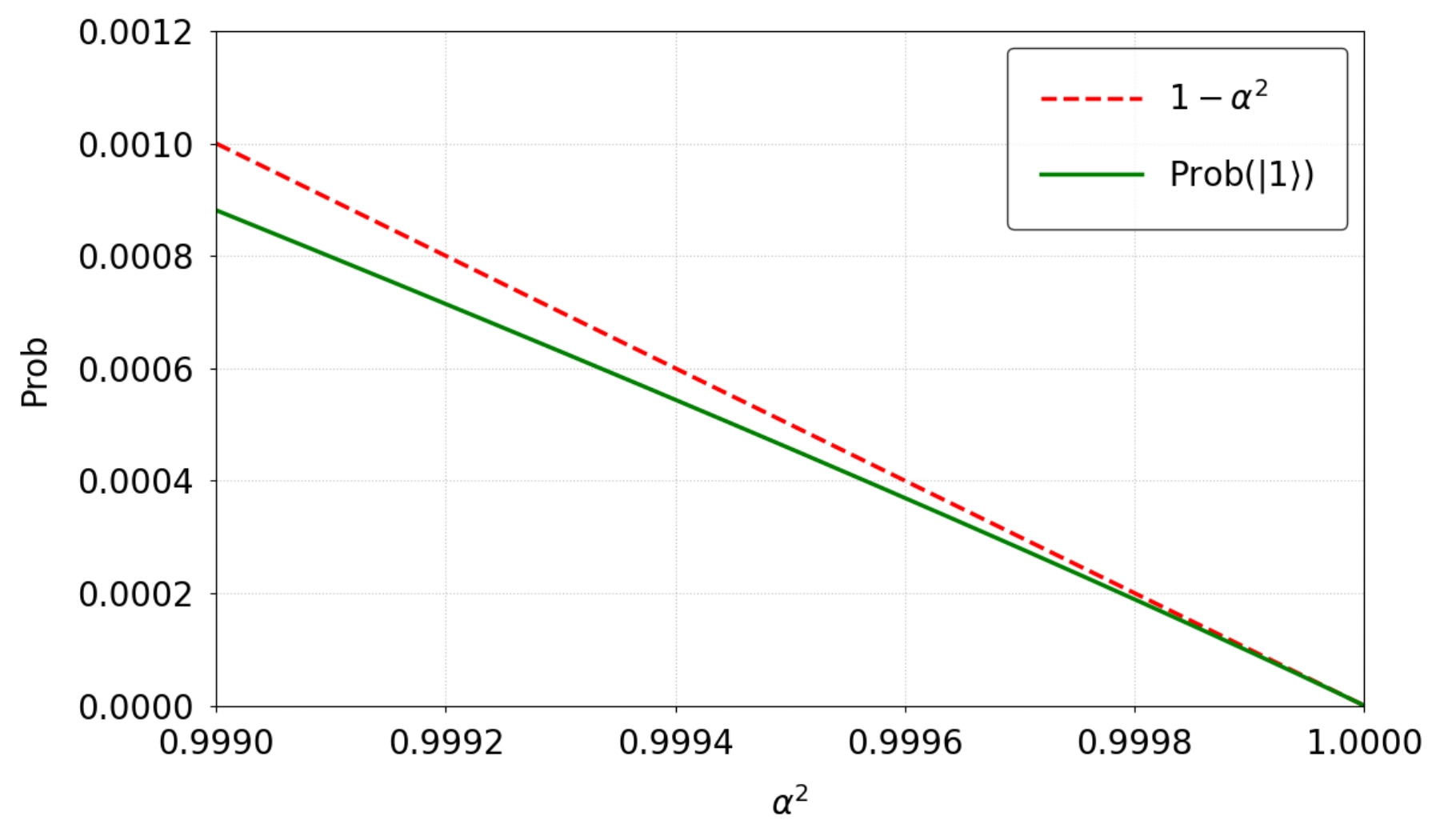}
  \caption{The comparsion of the incorrect result probability,
    $\text{Prob}(\ket{1})$, and the probability of the incorrect single
    Hadamard gate result $1-\alpha^2$.}
  \label{fig3}
\end{figure}

\begin{figure}
  \centering
  \includegraphics[width=0.8\columnwidth]{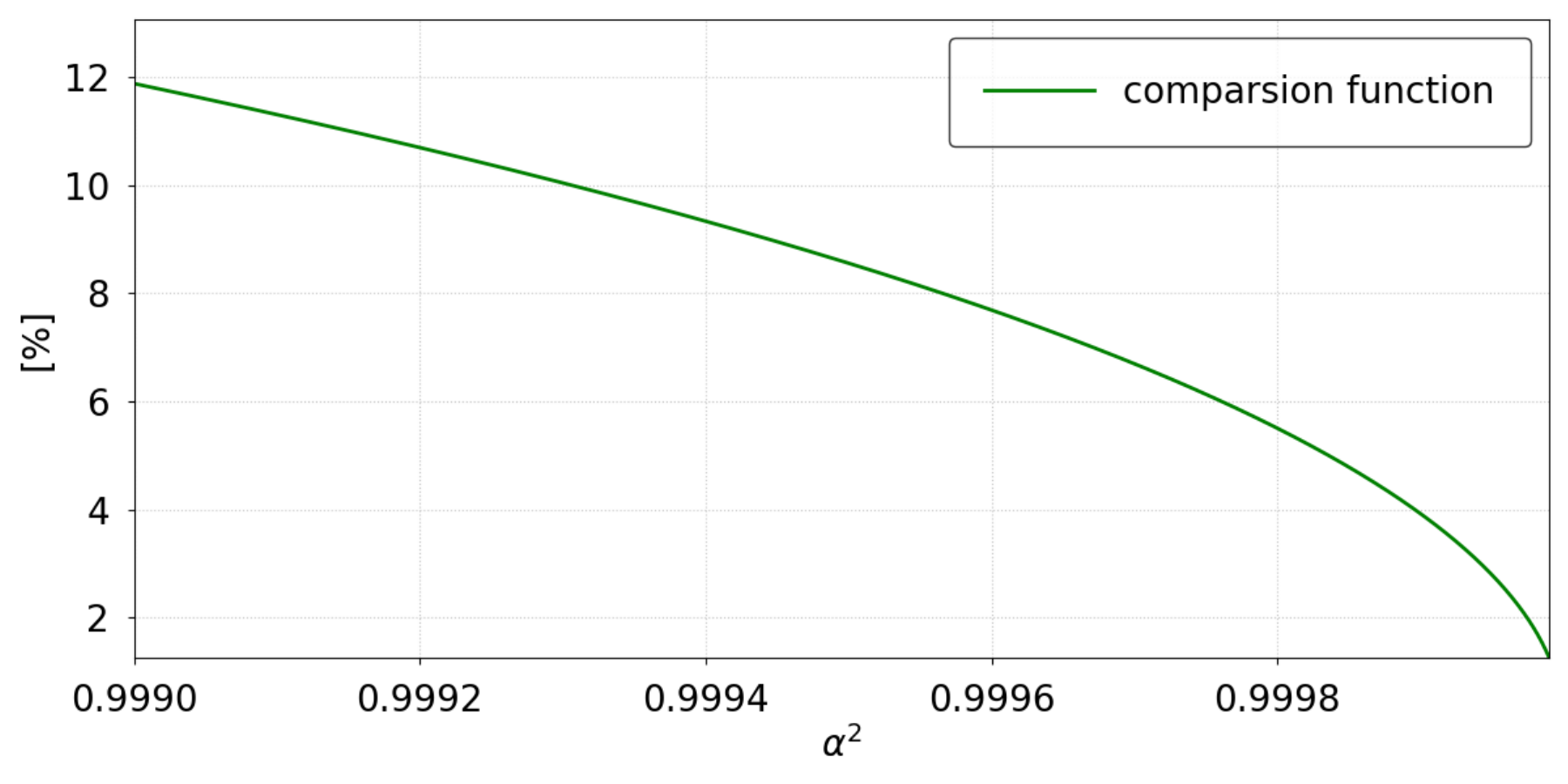}
  \caption{The ratio of the incorrect result probability
    $\text{Prob}(\ket{1})$ to the single Hadamard gate error
    $1-\alpha^2$.}
  \label{fig4}
\end{figure}

The coefficient $1-\alpha^2$ denotes the probability of an incorrect
single Hadamard gate result. In Fig.~\ref{fig3} one can see that the
algorithm error, $\mathrm{Prob}(\ket{1})$, is smaller than the single
Hadamard gate error. Fig.~\ref{fig4} shows by what percentage the error
function $\mathrm{Prob}(\ket{1})$ is smaller than the single Hadamard
gate error $1-\alpha^2$.

%%%%%%%%%%%%%%%%%%%%%%%%%%%%%%%%%%%%%%%%%%%%%%%%%%%%%%%%%%%%%%%%%%%%%%%%%%%%%%%%
%%%%%%%%%%%%%%%%%%%%%%%%%%%%%%%%%%%%%%%%%%%%%%%%%%%%%%%%%%%%%%%%%%%%%%%%%%%%%%%%

%%%%%%%%%%%%%%%%%%%%%%%%%%%%%%%%%%%%%%%%%%%%%%%%%%%%%%%%%%%%%%%%%%%%%%%%%%%%%%%%
%%%%%%%%%%%%%%%%%%%%%%%%%%%%%%%%%%%%%%%%%%%%%%%%%%%%%%%%%%%%%%%%%%%%%%%%%%%%%%%%

%%%%%%%%%%%%%%%%%%%%%%%%%%%%%%%%%%%%%%%%%%%%%%%%%%%%%%%%%%%%%%%%%%%%%%%%%%%%%%%%
%%%%%%%%%%%%%%%%%%%%%%%%%%%%%%%%%%%%%%%%%%%%%%%%%%%%%%%%%%%%%%%%%%%%%%%%%%%%%%%%

The PEv model distinguishes between the two types of quantum gates. Let's take
as an example two Hadamard gates $H_1$ and $H_3$ in a~raw followed by
a~measurement. In the case of unitary gates the density matrix $\rho$ will be
transformed into
\begin{equation}
  \rho \to 
  \frac
  {\EOp_M \EOp_{H_3} \EOp_{H_1} \rho 
   \EOp_{H_1}^\dagger \EOp_{H_3}^\dagger \EOp_M^\dagger}
  {\mathrm{Tr} \left( \EOp_M \EOp_{H_3} \EOp_{H_1} \rho 
   \EOp_{H_1}^\dagger \EOp_{H_3}^\dagger \EOp_M^\dagger \right)}.
\end{equation}
The gates produce a~superposition of correct and incorrect results and only
the final measurement determines the outcome. In the case of projection gates,
the transformation reads:
\begin{equation}
  \rho \to 
  \frac
  {\EOp_M \EOp_{M_3} \EOp_{H_3} \EOp_{M_1} \EOp_{H_1} \rho 
   \EOp_{H_1}^\dagger \EOp_{M_1}^\dagger 
   \EOp_{H_3}^\dagger \EOp_{M_3}^\dagger \EOp_M^\dagger}
 {\mathrm{Tr} \left( 
   \EOp_M \EOp_{M_3} \EOp_{H_3} \EOp_{M_1} \EOp_{H_1} \rho 
   \EOp_{H_1}^\dagger \EOp_{M_1}^\dagger 
   \EOp_{H_3}^\dagger \EOp_{M_3}^\dagger \EOp_M^\dagger \right)}.
\end{equation}
The gate $H_1$ measures its output state and sends one result, correct or
incorrect, to $H_3$ which sends its measured result to the measuring
apparatus. It is clearly seen that the schemes of both physical processes are
different, which may play a~role in the construction of the hardware.

In our discussion all the elements of the quantum circuit are present during
the whole computation. This is obviously needed for the algorithm to correctly
compute the result. In a~real quantum circuit, however, the spontaneous
decoherence of the states will introduce errors. The quantum gates will occupy
certain segments in space-time, with which the particle has to have a~non-zero
overlap. The gates will also loose accuracy with time and need to be
periodically reset. All of this will result in the temporal parts of their
wave functions not to be constant in time, in particular
$\alpha=\alpha(\tau;t)$, and may lead to delayed-choice effects \cite{M-Z} as
well as more complicated error propagation. These topics will be discussed in
our subsequent paper.

%%%%%%%%%%%%%%%%%%%%%%%%%%%%%%%%%%%%%%%%%%%%%%%%%%%%%%%%%%%%%%%%%%%%%%%%%%%%%%%%
\bibliographystyle{apsrev4-2}
\bibliography{Deutsch}
\end{document}